\documentclass[12pt,a4paper]{article}
\usepackage{amsmath,amssymb}
\usepackage{graphicx}
\begin{document}
\newcounter{Series}
\setcounter{Series}{2}
\title{
A Diagrammer's Note on 
Superconducting Fluctuation Transport 
for Beginners: \\
Supplement. Boltzmann Equation and Fermi-Liquid Theory}

\author{
O. Narikiyo
\footnote{
Department of Physics, 
Kyushu University, 
Fukuoka 812-8581, 
Japan}
}

\date{
(April 1, 2021)
}

\maketitle
\begin{abstract}
The effect of the collision term of the Boltzmann equation 
is discussed for the diagonal conductivity in the absence of magnetic field 
and the off-diagonal conductivity within the linear order of magnetic field. 
The consistency between the Boltzmann equation and the Fermi-liquid theory 
is confirmed. 
The electron-electron interaction is totally taken into account 
via the self-energy of the electron. 
The Umklappness is taken into account as the geometrical factor. 
The current-vertex correction in the fluctuation-exchange (FLEX) approximation 
violates these conventional schemes. 
In the Appendix the Tsuji formula, 
the geometrical formula for the Hall conductivity is discussed. 
\end{abstract}


\section{Introduction}

This Note is the Supplement to {\ttfamily arXiv:1112.1513} 
and {\ttfamily arXiv:1203.0127}. 
In the section 6 of the former and the footnote 4 of the latter, 
the exact treatment of the collision term of the Boltzmann equation (BE) is discussed 
and its consistency with the Fermi-liquid theory (FLT) is commented. 
In this Note the comment is expanded to be traceable. 
Especially it is stressed that 
the electron-electron interaction is totally taken into account 
via the self-energy of the electron.
The Umklappness\footnote{
In this note I use the term $\lq\lq$Umklappness"
instead of $\lq\lq$Umklapp scattering"
to stress the fact that it is nothing but the ambiguity 
of the momentum-conservation modulo reciprocal lattice vectors. } 
is taken into account as the geometrical factor. 
As a by-product it becomes clear how the fluctuation-exchange (FLEX) approximation 
violates these conventional schemes\footnote{
I have made other criticisms on the current-vertex correction in the FLEX approximation 
in {\ttfamily arXiv:1204.5300} and {\ttfamily arXiv:1301.5996}. }. 

Although the other Notes in the series are written to be self-contained, 
this Note skips the calculations to obtain the results cited. 

\section{Collision Term}

The linearized Boltzmann equation\footnote{
As discussed in the section 5 in {\ttfamily arXiv:1112.1513} 
the temperature gradient is incorporated by the substitution 
\begin{equation}
e {\bf E}' = e {\bf E} - \xi_{\bf p} {\nabla T \over T}, 
\nonumber 
\end{equation}
and the effect of the collision term can be analyzed in the same manner as in this Note. 
See Pikulin, Hou and Beenakker: Phys. Rev. B {\bf 84}, 035133 (2011). }
in static and uniform electromagnetic field\footnote{
The complication arising from inhomogeneity or time-dependence of electromagnetic field 
is beyond the scope of the series of Notes. 
See, for example, Kita: Prog. Theor. Phys. {\bf 123}, 581 (2010) for more general case. } 
is 
\begin{equation}
e{\bf E}\cdot{\bf v}_{\bf p} 
{ \partial f(\xi_{\bf p}) \over \partial \xi_{\bf p} } 
+ e \big( {\bf v}_{\bf p} \times {\bf H} \big)\cdot 
{ \partial g_{\bf p} \over \partial {\bf p} } 
= C_{\bf p}. 
\label{Bol} 
\end{equation}
The collision term $C_{\bf p}$ is given as 
\begin{equation}
C_{\bf p} = 
\sum_{\bf p'} \Big\{ C_{\bf pp'}g_{\bf p'} - C_{\bf p'p}g_{\bf p} \Big\} 
\equiv - \sum_{\bf p'} \big( \tau_{\rm tr}^{-1} \big)_{\bf pp'}g_{\bf p'}, 
\label{coll} 
\end{equation}
with 
\begin{equation}
\big( \tau_{\rm tr}^{-1} \big)_{\bf pp'} = 
{ 1 \over \tau_{\bf p} }\delta_{\bf p,p'} - C_{\bf pp'}, 
\end{equation}
and 
\begin{equation}
{ 1 \over \tau_{\bf p} } \equiv \sum_{\bf p'} C_{\bf p'p}. 
\end{equation}

It is evident in (2) 
that the scattered-in term and the scattered-out term are 
the same interaction process\footnote{
If the scattering is restricted on the Fermi sphere, 
all the scattering events are treated on equal footing completely 
and the collision term for the component $g_{lm}$ of spherical harmonics  $Y_{lm}$ 
becomes
\begin{equation}
-{1 \over \tau_l} g_{lm}, 
\nonumber 
\end{equation}
as (6.20) in \cite{Pei}. 
This result strongly suggests 
that the effect of the interaction is essentially taken into account by the life-time 
and the other have only geometrical effects. } 
but with different directions. 
More direct representation of this point is seen 
in the expression of the collision term in the quantum Boltzmann equation 
\begin{equation}
\Sigma^< G^> - \Sigma^> G^<,
\end{equation}
as (8.293) in \cite{Mah}. 
Namely, the effect of the interaction is totally expressed 
in terms of the self-energy $\Sigma$. 
The life-time $\tau_{\bf p}$ is determined by the imaginary part of the self-energy. 

Thus the current-vertex correction in the FLEX approximation seems to be out of control. 
It violates the scheme of BE 
where the collision term is expressed in terms of the self-energy. 
It also violates the scheme of FLT, 
because FLT and BE give the same result for the conductivity tensor 
as discussed in the following. 
The correct current-vertex correction should be expressed in terms of the self-energy. 

\section{Diagonal Conductivity}

The linearized Boltzmann equation is solved exactly~\cite{KSV} 
as discussed in the section 6 of {\ttfamily arXiv:1112.1513} 
and the conductivity tensor per spin is expressed as 
\begin{equation}
\sigma^{\mu\nu} = e^2 \sum_{\bf p'} \sum_{\bf p} 
v_{\bf p'}^\mu A_{\bf p'p}^{-1} v_{\bf p}^\nu 
\Big( - { \partial f(\xi_{\bf p}) \over \partial \xi_{\bf p} } \Big), 
\end{equation}
with 
\begin{equation}
A_{\bf pp'} = \big( \tau_{\rm tr}^{-1} \big)_{\bf pp'} 
- e \, \Big( {\bf v}_{\bf p} \times 
{ \partial \over \partial {\bf p} } \Big) 
\cdot {\bf H} \, \, \delta_{\bf pp'}. 
\end{equation}

\begin{figure}[htbp]
\begin{center}
\includegraphics[width=10.0cm]{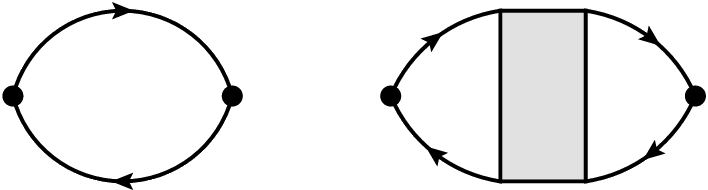}
\caption{
The particle-hole pair in the left diagram, 
which represents the skeleton contribution to $\sigma^{xx}$, 
results in $\tau_{\bf p}$. 
The interaction between particle-hole pair in the right diagram 
represents the infinite repetition of $C_{\bf pp'}$ and 
results in the renormalization of $\tau_{\bf p}$ into ${\tilde \tau}_{\bf p}$. 
Every dot at the ends of the diagrams represents $ev_{\bf p}^x$. 
}
\label{fig:Fig1}
\end{center}
\end{figure}

The diagonal conductivity in the absence of magnetic field is obtained as 
\begin{equation}
\sigma^{xx} = e^2 \sum_{\bf p'} \sum_{\bf p} 
v_{\bf p'}^x \big( \tau_{\rm tr} \big)_{\bf p'p} v_{\bf p}^x 
\Big( - { \partial f(\xi_{\bf p}) \over \partial \xi_{\bf p} } \Big). 
\label{sigma-xx} 
\end{equation}
Diagrammatically (\ref{sigma-xx}) is expressed as Fig.~\ref{fig:Fig1}. 
This result of BE is consistent with that of FLT with ${\bf k}=0$ and $\omega=0$ 
seen in (24) of \cite{Eli} with (27) and in (4.38) of \cite{YY}. 

\vskip 4mm

The expression (\ref{sigma-xx}) is not desirable, 
because it is not reduced into single-particle quantities. 
Such a reduction is the scheme of BE and 
can be accomplished easily in the isotropic case as   
\begin{equation}
\sigma^{xx} = e^2 \sum_{\bf p} 
v_{\bf p}^x {\tilde \tau}_{\bf p} v_{\bf p}^x 
\Big( - { \partial f(\xi_{\bf p}) \over \partial \xi_{\bf p} } \Big), 
\label{sigma-sym} 
\end{equation}
where ${\tilde \tau}_{\bf p}$ is a scalar. 
This result\footnote{
Such a symmetry in $v_{\bf p}^x$ is obvious in the memory-function formalism~\cite{HH}. } 
of BE is consistent with that of FLT 
seen in (6.23) of \cite{YY} 
which takes into account the effect of Umklappness properly. 
Here the electron-electron interaction is totally taken into account 
via the self-energy of the electron; 
${\bf v}_{\bf p}$ is renormalized\footnote{
Here the group velocity ${\bf v}_{\bf p}$ is determined 
by the renormalized band $\varepsilon({\bf p})$
of the quasi-particle as 
${\bf v}_{\bf p}=\partial\varepsilon({\bf p})/\partial{\bf p}$. } 
by the real part of the self-energy 
and ${\tilde \tau}_{\bf p}$ is determined\footnote{
We should evaluate an additional factor $C_{\bf p}$ 
resulted from the Umklappness so that 
${\tilde \tau}_{\bf p} = \tau_{\bf p} / C_{\bf p}$ 
as seen in (6.23) of \cite{YY}. 
It should be noted that the effects of interaction and Umklappness are separable. 
The former determines the life-time $\tau_{\bf p}$ 
and the latter determines the factor $C_{\bf p}$ 
which reflects the momentum-conservation with additional reciprocal lattice vectors. } 
by the imaginary part of the self-energy. 

The symmetric form of (\ref{sigma-sym}) is expected from 
the vector character of the current vertex $v_{\bf p}^x$; 
the none-zero contribution after the ${\bf p}$-summation 
comes from the pair of observed current $e{\bf v}_{\bf p}$ 
and the current $e{\bf v}_{\bf p}$ coupled to the electric field 
with the same momentum. 
In other words 
the cause and the effect are in the same direction in isotropic systems\footnote{
Even for anisotropic case, 
using the orthonormal vector set based on the Fermi-surface harmonics~\cite{All}, 
it is concluded that $\sigma^{xx}$ is proportional to $(e v_{\bf p}^x)^2$. }. 
Perturbationally 
both currents should be $e{\bf v}_{\bf p}$ 
at the vertices of the observation and the coupling to the electric field. 
In terms of BE 
the electric current is given by $e{\bf v}_{\bf p}g_{\bf p}$ 
where the first order deviation of the distribution function $g_{\bf p}$ 
is proportional to the strength of the perturbation $e{\bf v}_{\bf p}\cdot{\bf A}$ 
where ${\bf E}=-\partial{\bf A}/\partial t$. 
In any way the contribution to the conductivity $\sigma^{xx}$ 
is proportional to $(e v_{\bf p}^x)^2$. 

Even in anisotropic cases\footnote{
The symmetric form is obtained even for anisotropic cases; 
the conductivity is given as (1.4) in 
Okabe: J. Phys. Soc. Jpn. {\bf 68}, 2721 (1999) 
where $l_{\bf p}^x$ is decomposed into ${\tilde \tau}_{\bf p} v_{\bf p}^x$ 
as the equation between (4.3) and (4.4) in 
Okabe: J. Phys. Soc. Jpn. {\bf 67}, 4178 (1998). } 
the reduction can be accomplished symmetrically 
with the help of the Fermi-surface harmonics~\cite{All}\footnote{
On the basis of the Fermi-surface harmonics, 
all the elements in the theory are expressed 
as the polynomials of the velocity 
which is invariant under the transformation, 
${\bf p}\rightarrow{\bf p}+{\bf K}$, 
with ${\bf K}$ being a reciprocal lattice vector. 
Thus it becomes clear that the Umklappness is nothing but 
the condition for the summation over ${\bf p}$. } 
as seen in (2.1) of \cite{MF} 
which also takes into account the effect of Umklappness properly. 

On the other hand, the asymmetric expression\footnote{
By introducing the expansion of $J_{\bf p}^x$ 
in terms of the Fermi-surface harmonics, 
it is shown that all the terms orthogonal to $v_{\bf p}^x$ 
do not contribute to $\sigma^{xx}$ in Maebashi and Fukuyama: 
J. Phys. Soc. Jpn. {\bf 66}, 3577 (1997). } 
\begin{equation}
\sigma^{xx} = e^2 \sum_{\bf p} 
v_{\bf p}^x {1 \over 2\gamma_{\bf p}} J_{\bf p}^x 
\Big( - { \partial f(\xi_{\bf p}) \over \partial \xi_{\bf p} } \Big), 
\end{equation}
seen in (1.2) of \cite{KY} is misleading; 
it gives a wrong impression 
that there might be something in the current-vertex correction 
renormalizing $v_{\bf p}^x$ into $J_{\bf p}^x$. 
I suspect that the current-vertex correction in the FLEX approximation \cite{Kon} 
is driven by such a wrong impression. 
Although the vertex correction should be in harmony with 
the imaginary part of the single-particle self-energy, 
that in the FLEX approximation is out of control. 

\section{Off-diagonal Conductivity}

\begin{figure}[htbp]
\begin{center}
\includegraphics[width=12.6cm]{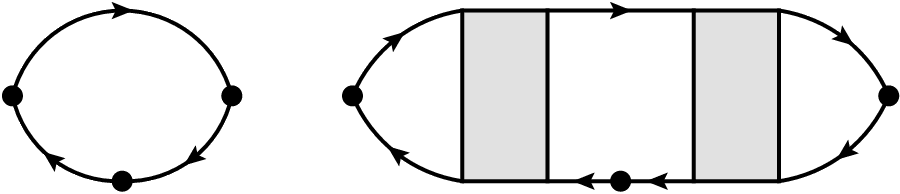}
\caption{
Here I show only two typical diagrams which contribute to $\sigma^{xy}$. 
The skeleton process in the left diagram results in $(\tau_{\bf p})^2$. 
The right diagram contains the skeleton structure at its center 
and the interactions at both ends result in 
the renormalization of $(\tau_{\bf p})^2$ into $({\tilde \tau}_{\bf p})^2$. 
Only the most divergent contribution in terms of $1/\tau_{\bf p}$ 
is taken into account in FLT 
and the skeleton structure is necessary for such a contribution. 
In each diagram the left-end and right-end dots represent
$ev_{\bf p}^x$ and $ev_{\bf p}^y$ respectively 
and the center dot represents the coupling $e{\bf v}_{\bf p}\cdot{\bf A}$ 
where ${\bf H} = i ({\bf k} \times {\bf A})$. } 
\label{fig:Fig2}
\end{center}
\end{figure}

Combined with the discussion leading to (\ref{sigma-sym}), 
the solution of BE for $\sigma^{xy}$ 
in the section 6 of {\ttfamily arXiv:1112.1513}
is reduced into 
\begin{equation}
\sigma^{xy} = e^3 \sum_{\bf p} 
v_{\bf p}^x {\tilde \tau}_{\bf p}\, 
\Big( {\bf v}_{\bf p} \times { \partial \over \partial {\bf p} } \Big) \cdot {\bf H}\, \, 
{\tilde \tau}_{\bf p} v_{\bf p}^y\, 
\Big( - { \partial f(\xi_{\bf p}) \over \partial \xi_{\bf p} } \Big), 
\label{sigma-xy} 
\end{equation}
within the linear order of magnetic field. 
Diagrammatically (\ref{sigma-xy}) is expressed as Fig.~\ref{fig:Fig2}. 
This result of BE is consistent with that of FLT\footnote{
Since the current-vertex correction there is not reduced into the single-particle quantities, 
(1.7) in \cite{KY} is misleading as discussed in the previous section. } 
seen in (35) of \cite{KF}\footnote{
An important point stressed in this reference is 
the absence of the interaction renormalization in the Hall coefficient. }. 
Within the linear response 
the collision term is evaluated in the absence of the external electromagnetic fields 
so that there is no new interaction effect by the introduction of the magnetic field. 
Namely, the introduction of the magnetic field is not a many-body problem 
but a single-body problem\footnote{
The effect of the Lorentz force on the single-body state is the problem. } 
as discussed in the next section. 

\vskip 4mm

This expression (\ref{sigma-xy}) is unsatisfactory by the following two reasons. 
(i) The Onsager relation, $\sigma^{yx}=-\sigma^{xy}$, is not evident. 
 (See {\ttfamily arXiv:2011.04421} for details.) 
(ii) The derivative of ${\tilde \tau}_{\bf p}$ is not found in the diagrammatic analysis. 
These are resolved as follows. 

I anti-symmetrize $\sigma^{xy}$ by force\footnote{
The same procedure is necessary 
to obtain a beautiful geometric formula of ${\hat \sigma}^{xy}$ 
in Ong: Phys. Rev. B {\bf 43}, 193 (1991) 
where the element of the line-integral ${\rm d}l_x l_y$ 
obtained from $\sigma^{xy}$ 
is anti-symmetrized into $({\rm d}{\vec l}\times{\vec l})_z$. } 
into 
${\hat \sigma}^{xy} \equiv (\sigma^{xy} - \sigma^{yx})/2$ where 
\begin{equation}
\sigma^{yx} = e^3 \sum_{\bf p} 
v_{\bf p}^y {\tilde \tau}_{\bf p}\, 
\Big( {\bf v}_{\bf p} \times { \partial \over \partial {\bf p} } \Big) \cdot {\bf H}\, \, 
{\tilde \tau}_{\bf p} v_{\bf p}^x\,  
\Big( - { \partial f(\xi_{\bf p}) \over \partial \xi_{\bf p} } \Big).
\end{equation}

If ${\bf H}=(0,0,H)$, we obtain 
\begin{equation}
{\hat \sigma}^{xy} = - {\hat \sigma}^{yx} 
= {e^3 H \over 2} \sum_{\bf p} 
{\tilde \tau}_{\bf p}^2\, 
h({\bf v}_{\bf p}, { \partial \over \partial {\bf p} })\, 
\Big( - { \partial f(\xi_{\bf p}) \over \partial \xi_{\bf p} } \Big),
\label{general} 
\end{equation}
with 
\begin{equation}
h({\bf v}_{\bf p}, { \partial \over \partial {\bf p} }) \equiv 
  (v^x_{\bf p})^2 { \partial v^y_{\bf p} \over \partial p^y } 
- v^x_{\bf p} v^y_{\bf p} { \partial v^y_{\bf p} \over \partial p^x } 
- v^y_{\bf p} v^x_{\bf p} { \partial v^x_{\bf p} \over \partial p^y } 
+ (v^y_{\bf p})^2 { \partial v^x_{\bf p} \over \partial p^x }. 
\end{equation}
This result of BE is consistent with the general result of FLT 
seen in (3.21) of \cite{KY}. 
(\ref{general}) reduces to (\ref{sigma-xy}) when $x$- and $y$- directions 
are equivalent as discussed in \cite{KY}.    

It should be noted that the derivatives of ${\tilde \tau}_{\bf p}$ cancel out 
through this anti-symmetrization. 
 
\section{Coupling to Magnetic Field}

The introduction of the magnetic field is discussed by various means. 
I will comment on three kinds of means in the following. 
All lead to the consistent result with BE\footnote{
In the derivation of BE via the Wigner function 
\begin{equation}
f_{\rm W}({\bf p},{\bf R},t) = \int {\rm d}{\bf r} 
\exp \big[ -i {\bf r}\cdot\big({\bf p} - e{\bf A}({\bf R},t)\big) \big] 
\Big\langle {\bf R}+{{\bf r}\over 2} \Big|
{\hat \rho}(t) 
\Big| {\bf R}-{{\bf r}\over 2} \Big\rangle,
\nonumber  
\end{equation}
the Lorentz force appears as seen in (4.78) of \cite{Ram}. }. 
It should be noted that the introduction of the magnetic field is a single-body problem. 

\vskip 10pt

\noindent
{\bf (A) Vector Potential}

\noindent
If we set ${\bf A}({\bf x}) = \exp(i{\bf k}\cdot{\bf x})$ 
with a constant vector ${\bf A}$,  
the magnetic field ${\bf H}$ is given by
\begin{equation}
{\bf H} = i ({\bf k} \times {\bf A}), 
\end{equation}
in the limit of ${\bf k}\rightarrow 0$.  
For anisotropic systems 
the extraction of this factor from Feynman diagrams with full interaction 
is rather complicated task\footnote{ 
This task can be circumvented by use of the Ward identity 
as discussed in Itoh: J. Phys. F {\bf 14}, L89 (1984). } 
as done in \cite{KY,FEW}. 
However, it is sufficient to show the means 
in the case of the relaxation-time approximation, 
because the interaction effect beyond this approximation 
only leads to the renormalization of $\tau_{\bf p}$ 
as has been discussed in the previous section. 
This task\footnote{
For the extraction of the factor, 
$(k^x A^y - k^y A^x)$, 
see Fig.\ 1 of {\ttfamily arXiv:1203.0127}. 
The processes (a) and (b) lead to the contribution proportional to 
\begin{equation}
  v^x_{\bf p} \big( v^x_{\bf p}A^x + v^y_{\bf p}A^y \big) 
\big( { \partial v^y_{\bf p} \over \partial p^x } k^x 
    + { \partial v^y_{\bf p} \over \partial p^y } k^y \big), 
\nonumber 
\end{equation}
where ${\bf A}$-linear term comes from the propagator 
and ${\bf k}$-linear term comes from the current, 
and the process (c) leads to 
\begin{equation}
- v^x_{\bf p} \big( v^x_{\bf p}k^x + v^y_{\bf p}k^y \big) 
\big( { \partial v^y_{\bf p} \over \partial p^x } A^x 
    + { \partial v^y_{\bf p} \over \partial p^y } A^y \big), 
\nonumber 
\end{equation}
where ${\bf k}$-linear term comes from the propagator 
and ${\bf A}$-linear term comes from the current,   
for general anisotropic case 
when ${\bf A}=(A^x,A^y,0)$ and ${\bf k}=(k^x,k^y,0)$. 
These results are obtained by the same manner 
as in the isotropic case, {\ttfamily arXiv:1203.0127}, 
only by generalizing the electric current (8) there to 
\begin{equation}
j_\mu^H(0) = e \sum_\sigma \sum_{\bf p} 
{ \partial \varepsilon({\bf p}-e{\bf A}) \over \partial p^\mu } 
c_{{\bf p}\sigma}^\dagger c_{{\bf p}\sigma}, 
\nonumber 
\end{equation}
and lead to (\ref{form}) in the next footnote 
times the magnetic field $H=i(k^x A^y - k^y A^x)$ where ${\bf H}=(0,0,H)$. 
Here the ${\bf k}$-linear expansion of $v_{\bf p}^y$ in (a) and (b) is generalized as 
\begin{equation}
{1 \over m} k^y \rightarrow 
{ \partial^2 \varepsilon({\bf p}) \over \partial p^x \partial p^y } k^x +  
{ \partial^2 \varepsilon({\bf p}) \over \partial p^y \partial p^y } k^y, 
\nonumber 
\end{equation}
where $\partial \varepsilon({\bf p}) / \partial {\bf p} = {\bf v}_{\bf p}$. 
The ${\bf A}$-linear expansion in (c) is generalized as 
\begin{equation}
{1 \over m} A^y \rightarrow 
{ \partial^2 \varepsilon({\bf p}) \over \partial p^x \partial p^y } A^x +  
{ \partial^2 \varepsilon({\bf p}) \over \partial p^y \partial p^y } A^y. 
\nonumber 
\end{equation}
} 
is a straightforward one. 

\vskip 10pt

\noindent
{\bf (B) Magnetic Flux}

\noindent
The effect of the magnetic field appears as 
the phase factor of the electron propagator as  
\begin{equation}
G({\bf r}_1,{\bf r}_2)
=\exp\big[ ie \Phi({\bf r}_1,{\bf r}_2) \big] G({\bf r}_1 - {\bf r}_2). 
\end{equation}
Namely, the interaction affects only 
translationally invariant propagator $G({\bf r}_1 - {\bf r}_2)$ 
and the phase $\Phi({\bf r}_1,{\bf r}_2)$ is determined geometrically. 
These two are separable and 
the propagator in the presence of the magnetic field 
$G({\bf r}_1,{\bf r}_2)$ 
is the product of these. 
The calculation of the loop diagram which corresponds to the conductivity 
leads to the magnetic flux 
as discussed in \cite{KF,Ede}. 

\vskip 10pt

\noindent
{\bf (C) Cyclotron Motion}

\noindent
The magnetic field enters into the equation of motion as the cyclotron frequency as 
\begin{equation}
{\dot J}_y = {1 \over i} \big[ J_y, K \big] = \omega_c J_x, 
\end{equation}
for isotropic case\footnote{
Even for anisotropic case 
the equation of motion for the center-of-mass current is easily calculated as 
\begin{equation}
{\dot J}_y = eH \big( m^{-1}_{yx} J_y - m^{-1}_{yy} J_x \big). 
\nonumber 
\end{equation}
Classically the Lorentz force 
\begin{equation}
{\hat m}{\dot {\bf v}} = e ( {\bf v} \times {\bf H}), 
\nonumber 
\end{equation}
leads to 
\begin{equation}
\left( \begin{array}{c} {\dot v}_x \\ {\dot v}_y \\ \end{array} \right) = eH 
\left( \begin{array}{cc} m^{-1}_{xx} & m^{-1}_{xy} \\ 
                   m^{-1}_{yx} & m^{-1}_{yy} \\ \end{array} \right) 
\left( \begin{array}{c} v_y \\ -v_x \\ \end{array} \right), 
\nonumber 
\end{equation}
when ${\bf H}=(0,0,H)$. 
Since $\sigma^{xy}$ is proportional to the expectation value of 
$ v_x {\dot v}_y $, we obtain the factor 
\begin{equation}
v_x \big( m^{-1}_{yy} v_x - m^{-1}_{yx} v_y \big), 
\nonumber 
\end{equation}
as seen in (12.5.9) of Ziman: {\it Electrons and Phonons} 
(Clarendon, Oxford, 1960). 
The general expression of this factor becomes 
\begin{equation}
v^x_{\bf p} \Big(
                    v^x_{\bf p} { \partial \over \partial p^y } 
                  - v^y_{\bf p} { \partial \over \partial p^x } 
\Big) v^y_{\bf p} 
= 
{ \partial \varepsilon({\bf p}) \over \partial p^x } \Big(   
{ \partial \varepsilon({\bf p}) \over \partial p^x }
{ \partial^2 \varepsilon({\bf p}) \over \partial p^y \partial p^y }  
- 
{ \partial \varepsilon({\bf p}) \over \partial p^y }   
{ \partial^2 \varepsilon({\bf p}) \over \partial p^x \partial p^y }  
\Big). 
\label{form} 
\end{equation}
}
where ${\bf J}$ is the center-of-mass current\footnote{
The equation of motion for the center-of-mass current 
is independent of the interaction so that the factor 
$ v_x ( m^{-1}_{yy} v_x - m^{-1}_{yx} v_y )$ 
is independent of the interaction. 
Such a conclusion is a corollary of Kohn's theorem: 
Phys. Rev. {\bf 123}, 1242 (1961). }. 
The calculation of the memory function\footnote{
The scalar memory function is insufficient to construct a consistent transport theory. 
We have to use the matrix memory function as discussed in \cite{HH,MF}. } 
leads to the consistent result~\cite{TYQ,SI} with (\ref{sigma-xy}). 

\section{Remarks}

Finally it is stressed that the effect of 
the electron-electron interaction on the conductivity tensor 
is totally taken into account via the self-energy of the electron. 
It is completely embodied in the conventional schemes of BE and FLT 
where the Umklappness and the anisotropy of the Fermi surface 
are properly taken into account. 

The sections of {\bf Exercise} and {\bf Acknowledgements} 
are common to those in {\ttfamily arXiv:1112.1513} so that I do not repeat here. 
Eq.\ (76) in {\ttfamily arXiv:1112.1513} should be   
\begin{equation}
\left( \begin{array}{c} {\bf J}^e \\ {\bf J}^Q \\ \end{array} \right) = 
\left( \begin{array}{cc} \sigma & \alpha \\ 
                   \tilde\alpha & \kappa \\ \end{array} \right) 
\left( \begin{array}{c} {\bf E} \\ -\nabla T \\ \end{array} \right). 
\nonumber 
\end{equation}

\section{Appendix}

In this Appendix\footnote{
This Appendix is the reproduction of http://hdl.handle.net/2324/2344834.} 
the Tsuji formula for the Hall conductivity in metals is discussed 
in Haldane's framework. (See {\ttfamily arXiv:2011.04421} for details.)

The Tsuji formula [Tsuji] is widely known as a geometrical formula 
for the Hall conductivity in metals under weak magnetic field. 
Since it was derived under the assumption of the cubic symmetry, 
Haldane [Haldane] tried to eliminate the assumption. 
Here we discuss the Tsuji formula using Haldane's framework. 
However, our conclusion is different from Haldane's. 
The details\footnote{
This Appendix is a nutshell of our previous note, http://hdl.handle.net/2324/1957531.}
are described in http://hdl.handle.net/2324/1957531. 

In usual notation the weak-field DC Hall conductivity tensor $\sigma^{xy}$ per spin 
is given by [Tsuji, Haldane] 
\begin{equation}
\sigma^{xy} 
= e^3 B \int { {\rm d}S \over (2\pi)^3 } 
\left( v^x \!,\, v^y \right)
\begin{pmatrix}
M^{-1}_{yy} & -M^{-1}_{yx} \\
0 & 0
\end{pmatrix}
\begin{pmatrix}
v^x \\
v^y 
\end{pmatrix}
{ \tau^2 \over |{\vec v}| }, 
\label{1}
\end{equation}
for the Fermi surface contribution in metals. 
Throughout this note we only consider the contribution from a single sheet of the Fermi surface. 
Here the magnetic field is chosen as ${\vec B}=(0,0,B)$. 
The quasi-particle velocity ${\vec v}=(v^x,v^y,v^z)$ 
and the effective mass tensor $M_{\alpha\beta}$ 
are given by the derivative of the quasi-particle energy $\varepsilon$: 
$v^\alpha = \partial \varepsilon / \partial k^\alpha$ and 
$M_{\alpha\beta}^{-1} = \partial^2 \varepsilon / \partial k^\alpha \partial k^\beta$. 
Since the contribution of the derivative of $\tau$ does not appear in the antisymmetric tensor 
$(\sigma^{xy} - \sigma^{yx})/2$, we have dropped it. 

Experimentally $\sigma^{xy}$ is obtained from the measurement 
where we measure the current in $x$-direction 
under the electric field in $y$-direction and the magnetic field in $z$-direction. 
If we measure the current in $y$-direction 
under the electric field in $x$-direction and the magnetic field in $z$-direction, 
we obtain $\sigma^{yx}$ described as 
\begin{equation}
-\sigma^{yx} 
= e^3 B \int { {\rm d}S \over (2\pi)^3 } 
\left( v^x \!,\, v^y \right)
\begin{pmatrix}
0 & 0 \\
-M^{-1}_{xy} & M^{-1}_{xx}
\end{pmatrix}
\begin{pmatrix}
v^x \\
v^y 
\end{pmatrix}
{ \tau^2 \over |{\vec v}| }.
\label{2}
\end{equation}

Haldane [Haldane] introduced the symmetric tensor 
$e^3 B \gamma_{zz} \equiv (\sigma^{xy} - \sigma^{yx})/2$. 
Eq. (\ref{1}) and Eq. (\ref{2}) lead to 
\begin{equation}
\gamma_{zz} 
= {1 \over 2} \int { {\rm d}S \over (2\pi)^3 } 
\left( v^x \!,\, v^y \right)
\begin{pmatrix}
M^{-1}_{yy} & -M^{-1}_{yx} \\
-M^{-1}_{xy} & M^{-1}_{xx}
\end{pmatrix}
\begin{pmatrix}
v^x \\
v^y 
\end{pmatrix}
{ \tau^2 \over |{\vec v}| }. 
\label{3}
\end{equation}
Other symmetric tensors are introduced in the same manner as 
$e^3 B \gamma_{xx} \equiv (\sigma^{yz} - \sigma^{zy})/2$ and 
$e^3 B \gamma_{yy} \equiv (\sigma^{zx} - \sigma^{xz})/2$. 
As shown in the following the geometrical nature is captured by these symmetric tensors. 
It should be noted that our result, Eq. (\ref{3}), is different form Haldane's [Haldane]. 
The difference arises from the following fact. 
While Eq. (\ref{3}) contains $(\partial v^x / \partial k^y)/|{\vec v}|$, 
Haldane erroneously uses $\partial(v^x/|{\vec v}|) / \partial k^y$ instead. 

The target of our geometrical description is the mean curvature $H$ of the Fermi surface. 
It is given by 
\begin{eqnarray}
2H &=&
{1 \over |{\vec v}|^3 } \cdot
\Big[
\varepsilon_x \varepsilon_x ( \varepsilon_{yy} + \varepsilon_{zz} ) +
\varepsilon_y \varepsilon_y ( \varepsilon_{zz} + \varepsilon_{xx} ) + 
\varepsilon_z \varepsilon_z ( \varepsilon_{xx} + \varepsilon_{yy} ) \nonumber \\
&-&
\varepsilon_x ( \varepsilon_y \varepsilon_{yx} + \varepsilon_z \varepsilon_{zx} ) -
\varepsilon_y ( \varepsilon_x \varepsilon_{xy} + \varepsilon_z \varepsilon_{zy} ) -
\varepsilon_z ( \varepsilon_x \varepsilon_{xz} + \varepsilon_y \varepsilon_{yz} ) 
\Big],
\nonumber
\end{eqnarray} 
for any shape of the Fermi surface. 
Here we have used the notations $\varepsilon_\alpha \equiv v^\alpha$ 
and $\varepsilon_{\alpha\beta} \equiv M_{\alpha\beta}^{-1}$. 
 
The geometrical information in our master equation, Eq. (\ref{3}), is represented by $h_{zz}$ as 
\begin{equation}
\gamma_{zz} 
= \int { {\rm d}S \over (2\pi)^3 } 
h_{zz} \tau^2, 
\nonumber
\end{equation}
with 
\begin{equation}
h_{zz} = {1 \over2 |{\vec v}| }
\left( 
  \varepsilon_x \varepsilon_x \varepsilon_{yy}
+ \varepsilon_y \varepsilon_y \varepsilon_{xx}
- \varepsilon_x \varepsilon_y \varepsilon_{yx}
- \varepsilon_y \varepsilon_x \varepsilon_{xy}
\right).
\nonumber
\end{equation}
Using 
\begin{equation}
h_{xx} = {1 \over2 |{\vec v}| }
\left( 
  \varepsilon_y \varepsilon_y \varepsilon_{zz}
+ \varepsilon_z \varepsilon_z \varepsilon_{yy}
- \varepsilon_y \varepsilon_z \varepsilon_{zy}
- \varepsilon_z \varepsilon_y \varepsilon_{yz}
\right),
\nonumber
\end{equation}
and
\begin{equation}
h_{yy} = {1 \over2 |{\vec v}| }
\left( 
  \varepsilon_z \varepsilon_z \varepsilon_{xx}
+ \varepsilon_x \varepsilon_x \varepsilon_{zz}
- \varepsilon_z \varepsilon_x \varepsilon_{xz}
- \varepsilon_x \varepsilon_z \varepsilon_{zx}
\right),
\nonumber
\end{equation}
additionally, we obtain 
\begin{equation}
\gamma_{zz} + \gamma_{xx} + \gamma_{yy} 
= \int { {\rm d}S \over (2\pi)^3 } H l^2, 
\label{4}
\end{equation}
with $l^2=|{\vec v}|^2\tau^2$. 
Our result, Eq. (\ref{4}), is applicable to any shape of the Fermi surface. 
In the case of cubic symmetry 
Eq. (\ref{4}) is reduced to the Tsuji formula [Tsuji, Haldane]  
\begin{eqnarray}
\gamma_{zz} = \gamma_{xx} = \gamma_{yy}
= \int { {\rm d}S \over (2\pi)^3 } { H \over 3 } l^2. 
\nonumber
\end{eqnarray}

Experimentally $\gamma_{cc}$ is obtained from the measurements of $\sigma^{ab}$ and $\sigma^{ba}$ 
where $(c,a,b)=(z,x,y),(x,y,z),(y,z,x)$. 
By summing six experimental results with different configurations we can use Eq. (\ref{4}). 

\vskip 12pt

[Tsuji] Tsuji: J. Phys. Soc. Jpn. {\bf 13}, 979 (1958). 

\vskip 6pt

[Haldane] Haldane: arXiv:cond-mat/0504227v2. 



\begin{thebibliography}{99}
\bibitem{xxx} 
As mentioned in the Introduction, 
this Note skips the calculations to obtain the results cited. 
Thus the character of the following references is different 
from those of the other Notes in this series. 
I only list the references convenient for my explanation. 
Neither originality nor priority is considered here. 

\bibitem{Pei}
Peierls: 
{\it Quantum Theory of Solids} 
(Clarendon, Oxford, 1955). 

\bibitem{Mah}
Mahan: 
{\it Many-Particle Physics} 3rd 
(Kluwer/Plenum, New York, 2000). 

\bibitem{KSV} 
Kotliar, Sengupta and Varma: 
Phys. Rev. B {\bf 53}, 3573 (1996). 

\bibitem{Eli} 
\'Eliashberg: 
Sov. Phys. JETP {\bf 14}, 886 (1962). 

\bibitem{YY} 
Yamada and Yosida: 
Prog. Theor. Phys. {\bf 76}, 621 (1986).

\bibitem{HH}
Hartnoll and Hofman: 
Phys. Rev. Lett. {\bf 108}, 241601 (2012). 

\bibitem{All} 
Allen: 
Phys. Rev. B {\bf 13}, 1416 (1976). 

\bibitem{MF} 
Maebashi and Fukuyama: 
J. Phys. Soc. Jpn. {\bf 67}, 242 (1998). 

\bibitem{KY}
Kohno and Yamada: 
Prog. Theor. Phys. {\bf 80}, 623 (1988). 

\bibitem{Kon} 
Kontani: 
Pep. Prog. Phys. {\bf 71}, 026501 (2008). 

\bibitem{KF} 
Khodas and Finkel'stein: 
Phys. Rev. B {\bf 68}, 155114 (2003). 

\bibitem{Ram}
Rammer: 
{\it Quantum Transport Theory} 
(Perseus Books, Reading Massachusetts, 1998). 

\bibitem{FEW} 
Fukuyama, Ebisawa and Wada: 
Prog. Theor. Phys. {\bf 42}, 494 (1969). 

\bibitem{Ede} 
Edelstein: 
JETP Lett. {\bf 67}, 159 (1998). 

\bibitem{TYQ} 
Ting, Ying and Quinn: 
Phys. Rev. B {\bf 16}, 5394 (1977). 

\bibitem{SI} 
Shiwa and Ishihara: 
J. Phys. C {\bf 16}, 4853 (1983). 

\end{thebibliography}
\end{document}